\documentclass[showpacs,preprintnumbers,amsmath,amssymb,eqsecnum]{revtex4}
\usepackage{graphicx}
\usepackage{subfigure}
\usepackage{epstopdf}
\usepackage{bm}
\usepackage{hyperref}
\def\cc{{\cal C}}

\begin{document}

\title{I-Love relation for incompressible stars and realistic stars}

\author{T.~K. Chan\footnote{Present address: Department of Physics,
University of California at San Diego, 9500 Gilman Drive, La
Jolla, CA 92093, USA. Email address: chantsangkeung@gmail.com},
AtMa P.~O. Chan\footnote{Present address: Department of Physics,
University of Illinois at Urbana-Champaign,
 Urbana, IL 61801-3080, USA. Email address: atma.pochan@gmail.com}}
\author{P.~T. Leung\footnote{Email:
ptleung@phy.cuhk.edu.hk}}
\affiliation{Physics Department and
Institute of Theoretical Physics, The Chinese University of Hong
Kong, Shatin, Hong Kong SAR, China.}

\date{\today}
\begin{abstract}
In spite of the diversity in the equations of state of nuclear
matter, the recently discovered I-Love-Q relations [Yagi and
Yunes, Science {\bf  341}, 365 (2013)], which relate the moment of
inertia, tidal Love number (deformability) and the spin-induced
quadrupole moment of compact stars, hold for various kinds of
realistic neutron stars and quark stars. While the physical origin
of such universality is still a current issue, the observation
that the I-Love-Q relations of incompressible stars can well
approximate those of realistic compact stars hints at a new
direction to approach the problem. In this paper, by establishing
recursive post-Minkowskian expansion for the moment of inertia and
the tidal deformability of incompressible stars, we analytically
derive the I-Love relation for incompressible stars and show that
the so obtained formula
 can be used to accurately predict the behavior of
realistic compact stars from the Newtonian limit to the maximum
mass limit.

\end{abstract}
 \pacs{04.40.Dg, 04.25.Nx, 97.60.Gb, 95.30.Sf}
\maketitle

\section{Introduction}
Recently, \citet{Yagi:2013long,Yagi:2013} have discovered the so
called ``I-Love-Q universal relations" prevailing in compact
stars, including both neutron stars (NSs) or quark stars (QSs). In
such relations, the moment of inertia $I$, the quadrupole tidal
Love number $\lambda_{ }$ (or, more precisely, tidal deformability
\citep{Damour:09p084035,Yagi_14_PRD}), and the spin-induced
quadrupole moment  $Q$ of compact stars are expressed in terms of
one another, with the stellar mass $M$ playing the role of a
scaling parameter. Soon after the discovery of
\citet{Yagi:2013long,Yagi:2013}, the I-Love-Q  relations were
generalized to several other cases, including binary systems with
strong dynamical tidal field \citep{Maselli:2013}, magnetized NSs
with sufficiently high rotation rates \citep{Haskell_14_mnras},
rapidly-rotating stars \citep{Pappas_14_prl,Chak_14_prl}, and
higher-order multipole moments induced by either tidal forces or
rotation \citep{Yagi_14_PRD,Yagi_hair_GR,Stein_hair_apj}.

These  relations are useful for several reasons
\citep{Yagi:2013long,Yagi:2013}. First,  they provide a link
directly connecting the I-Love-Q triplet. Once the mass of a
compact star is known, each one of of $I$, $\lambda_{}$ and $Q$
can lead to the determination of
 the other two. Second, in the
analysis of gravitational wave signals emitted during the late
stages of NS-NS binary mergers, they can break the degeneracy
between the  contributions due to the quadrupole moment and the
spin and hence enable more accurate measurement of the averaged
spin of the system
\citep{Flanagan:08p021502,Hinderer:08p1216,Yagi:2013long,Yagi:2013}.
Third, they can  identify the validity of other modified gravity
theories such as the Chern-Simons gravity and the
Eddington-inspired Born-Infeld (EiBI) gravity
\citep{Yagi:2013long,Yagi:2013,Sham_14_ApJ,Pani_Berti2014}.

On the other hand, the emergence of the I-Love-Q  relations is
also interesting from theoretical point of view. As is well known,
the physical characteristics of NSs (or QSs), including  mass,
radius, and moment of inertia, are usually obscured by various
uncertainties in the equation of state (EOS) of nuclear matter (or
quark matter). In fact, nuclear physicists have been using
different characteristics (e.g., the mass-radius relation, the
maximum mass and gravitational wave spectrum)  of NSs (or QSs) as
the test bed of various EOSs of dense matter (see, e.g.,
\citep{Andersson1996,Andersson1998,Lattimer:2001,Lattimer:2005p7082,lattimer2007nso,Ozel_09_PRD,2013arXiv1303.3282H,Latt_14_EPJ,Lattimer_2013_ApJ}).
Nonetheless, the I-Love-Q  relations are shown to hold for
different commonly accepted EOSs for both NSs and QSs
\citep{Yagi:2013long,Yagi:2013,Yagi_14_PRD}. What is the physical
origin of such universal behavior? \citet{Yagi:2013long,Yagi:2013}
have suggested two possible reasons for such universality: (i) the
relations are mainly dominated by  the low-density matter lying in
an layer between $70 \%$ and $90 \%$ of  the stellar radius where
the EOS
 is well known; and (ii) NSs merely follow the behavior of black-holes and
 obey the no-hair theorem in the high-compactness limit. However, more
recent investigation showed  that the I-Love-Q relations are
dominated by a thicker layer which is bounded between $50 \%$ and
$90 \%$ of the radius and comprises both high and low-density
matter \citep{whyI}. As a result, it is rather unlikely that the
similarity of EOSs in the low-density regime can give rise to the
I-Love-Q universality.

Moreover, it is interesting to note that  the I-Love-Q relations
are valid for incompressible stars and QSs
\citep{Yagi_14_PRD,ILoveQ_1}, which are constructed from EOSs
completely different from those of normal NSs. In particular,
\citet{ILoveQ_1} have attributed the I-Love-Q universality to (i)
the high stiffness of nuclear matter and quark matter, and (ii)
the I-Love-Q relations are stationary with respect to variations
of stiffness of stellar matter about the incompressible limit. As
a result, except very close to the low mass limit of NSs, where
NSs are made of relatively soft low-density substance, the
I-Love-Q relations of realistic NSs are well approximated by those
of incompressible stars to less than a few percent.

The main objective of the present paper is to derive analytically
the I-Love relation for incompressible stars. As discovered by
\citet{ILoveQ_1} and mentioned above, the relation so obtained is
also applicable to realistic NSs. We note that, to our knowledge,
so far the general relativistic I-Love-Q relations for realistic
NSs are only expressed in terms of the empirical formulas
suggested by \citet{Yagi:2013long,Yagi:2013} as follows:
\begin{equation}
\ln y_i = a_i + b_i \ln x_i + c_i \left( \ln x_i \right)^2 + d_i
\left(\ln x_i \right)^3 + e_i \left( \ln x_i \right)^4 ,
\label{eq:ILoveQ_fit}
\end{equation}
where $x_i$ and $y_i$ are any two of the I-Love-Q triplet, $a_i$,
$b_i$, $c_i$, $d_i$, and $e_i$ are some fitting coefficients (see
\citep{Yagi:2013long,Yagi:2013} for the values  of the these
coefficients). By establishing  recursive post-Minkowskian
perturbative schemes for the (scaled) moment of inertia and tidal
Love number (deformability) of incompressible stars, we obtain
simple yet accurate formulas expressing these two physical
quantities as functions (e.g., power series and fractions) of the
stellar compactness. After eliminating the stellar compactness
from the formulas, we find the I-Love relation for incompressible
stars, which is expressible in terms of simple algebraic functions
(see (\ref{incomilove_ds})  and (\ref{incomlovei_Pade})). Most
importantly, we show that the so derived I-Love relation is also
valid for realistic compact stars (including NSs and QSs) and
applies to both relativistic and Newtonian stars with high degree
of accuracy.

The paper is organized as follows. In Section \ref{TOV} we briefly
review the solution of the Tolman-Oppenheimer-Volkov (TOV)
equations \cite{Oppenheimer, Tolman}, which  govern the
hydrostatic equilibrium of relativistic stars, for incompressible
stars \citep{Tolman_book} and expand relevant  physical quantities
(e.g., pressure and metric coefficients) in power series of
compactness. In Sections \ref{moment} and \ref{Love}  we formulate
recursive perturbative expansions for the moment of inertia and
the tidal Love number (deformability) for incompressible stars,
respectively. In Section \ref{I_Love} we find analytic formulas
(\ref{incomilove_ds})  and (\ref{incomlovei_Pade})  relating the
moment of inertia and the Love number for incompressible stars and
show that these formulas also work nicely for realistic NSs and
QSs. We conclude our paper in Section \ref{Conclusion} with some
discussions. Unless otherwise stated explicitly, geometric units
in which $G=c=1$ are adopted.

\section{Hydrostatic Equilibrium of incompressible stars}\label{TOV}
Here we briefly review the hydrostatic equilibrium solution of
 a relativistic, non-rotating compact star made of  perfect fluid with a constant density $\rho_c$. Since the system
is stationary and spherically symmetric, the spacetime metric
admits the simple form
\begin{equation}\label{StaticMetric}
ds^{2}=-e^{\nu}dt^{2}+e^{\lambda}dr^{2}+r^{2}\left(d\theta^{2}+\sin^{2}\theta
d{ \phi}^{2}\right)~,
\end{equation}
where $t$, $r$, $\theta$, ${ \phi}$ are the standard Schwarzschild
coordinates, $\nu$ and $\lambda$ are some functions of $r$ (see,
e.g., \cite{Hartle, ShapiroComstar}). In particular, the metric
outside the star is given by the Schwarzschild metric with
$e^{\nu}=e^{-\lambda}=1-2M/r$, where $M$ is the total mass of the
star. Inside the star, the   TOV equations govern the variation of
the pressure $p$ and the metric coefficient $e^{\nu}$
\cite{Oppenheimer, Tolman}:
\begin{eqnarray}
\frac{dp}{dr}&=&-\frac{(m+4\pi r^{3}p)(\rho_c+p)}{r^{2}(1-2m/r)}~,\label{TOV1}\\
\frac{d\nu}{dr}&=&-\frac{2}{\rho+p}\frac{dp}{dr}~,\label{TOV2}
\end{eqnarray}
where  $m(r)= 4\pi\rho_c r^3/3$ is just the gravitational mass
enclosed within radius $r$ \cite{Oppenheimer, Tolman}. The other
metric coefficient $e^{\lambda}$ can be determined from $m(r)$ by
$e^{-\lambda}=1-2m(r)/r$. For an incompressible star with a given
radius $R$, the TOV equations together with the two boundary
conditions (i) $p(R)=0$ and (ii) $e^{\nu(R)}=1-2M/R$ completely
specify both $p(r)$  and $e^{\nu(r)}$ inside the star.

The analytical solution of the TOV equations, (\ref{TOV1}) and
(\ref{TOV2}), for incompressible stars, which is known as the
Schwarzchild constant-density interior solution
\citep{Tolman_book}, is given by:
\begin{eqnarray}
   p(r)&=&  \rho_c
   \left(\frac{\sqrt{1-2Mr^2/R^3}-\sqrt{1-2M/R}}{3\sqrt{1-2M/R}-\sqrt{1-2Mr^2/R^3}}\right), \label{interior_1} \\
e^{\nu(r)}&=&
\left(\frac{3}{2}\sqrt{1-\frac{2M}{R}}-\frac{1}{2}\sqrt{1-\frac{2M
r^2}{R^3}}\right)^2, \label{interior_2}
\end{eqnarray}
where $r \le R$. In order to keep the pressure positive and hence
physically acceptable, the compactness $\cc \equiv M/R$ of
incompressible stars must be less than $4/9$.

It can be readily shown from (\ref{interior_1}) and
(\ref{interior_2}) that both $p(r)$ and $e^{\nu(r)}$ can be
considered as functions of the scaled radius $x \equiv r/R$ and
the compactness $\cc$. Both of them can be expanded as Taylor
series in $\cc$,
\begin{eqnarray}
p(x;\cc)&=&p_{0}(x)+p_{1}(x)\cc+p_{2}(x)\cc^{2}+\cdots~,\\
e^{\nu}(x;\cc)& = & (e^{\nu})_0(x)  +  (e^{\nu})_1(x) \cc+
(e^{\nu})_2(x) \cc^2 + \cdots~,
\end{eqnarray}
where $p_{n}(x)$ and $(e^{\nu})_n(x)$, $n=0,1,2,\cdots$, are
functions of $x$ only. The leading expansions are given explicitly
by
\begin{eqnarray}
p(x;\cc)&=&\frac{1}{2}\rho_c\left(1-x^{2}\right)\cc
+\rho_c\left(1-x^{2}\right)\cc^{2} +\cdots~,\label{expp}\\
e^{\nu}(x;\cc)&=&1-(3-x^2)\cc
+\frac{3}{4}\left(1-x^{2}\right)^{2}\cc^{2}+\cdots~.\label{expnu}
\end{eqnarray}
For reference and illustration, we tabulate $\bar{p}_n (x) \equiv
p_n(x)/\rho_c$ and $(e^{\nu})_n (x)$ for  $0 \le n \le 6$ in
Tables~\ref{TSS2} and \ref{TSS3}. On the other hand, it is
straightforward to show that
\begin{eqnarray}
e^\lambda(x;\cc)&=&\frac{1}{1-2\cc x^2},\nonumber \\
&=&1+2\cc x^2+(2\cc x^2)^2 +\cdots~.\label{explamb}
\end{eqnarray}
In the following discussion, we will use the post-Minkowsian
expansions in (\ref{expp}), (\ref{expnu}) and (\ref{explamb}) as
the input to evaluate the moment of inertia and the tidal Love
number (deformability) of incompressible stars.
\section{Moment of Inertia}\label{moment}
 From
the analytic stellar profile derived above, in this section we
formulate a post-Minkowskian recursive perturbation scheme to
calculate the moment of inertia $I$ of incompressible stars in the
slowly rotating limit \cite{Hartle_1967,Hartle_1968,ComStar}.
Consider a uniformly rotating  star with a unit angular velocity.
Let $\Lambda$ be the angular velocity of the local inertial frame
due to the frame-dragging effect of the rotating star
\cite{Hartle_1967,Hartle_1968}. It satisfies the differential
equation \cite{Hartle_1967,Hartle_1968}:
\begin{equation}\label{moi1}
\frac{d}{dx}\Big(x^{4}j\frac{d\Lambda}{dx}\Big)+4x^{3}\frac{dj}{dx}(\Lambda-1)=0~,
\end{equation}
where $j(x)=e^{-(\lambda+\nu)/2}$ for $0 \le x \le 1$ and $j(x)=1$
for $x>1$. Outside the rotating star, the above equation can be
readily integrated to yield the result $\Lambda=2I/r^3$
\cite{Hartle_1967,Hartle_1968}. In particular, the surface value
of $\Lambda$, $\hat{\Lambda} \equiv \Lambda(x=1)$, is equal to
$2I/R^3$. Thus, the geometric factor  $a \equiv I/(MR^{2})$ is
given by $a=\hat{\Lambda}/2\cc$. It is obvious that for
incompressible stars $a=2/5$ in the Newtonian limit. However,
general relativistic effect could lead to modification in the
value of $a$.

To solve (\ref{moi1}) inside the star, similar to what we have
done in Section~\ref{TOV}, we expand $\Lambda(x)$ and $j(x)$ in
power series of $\cc$,
\begin{eqnarray}
\Lambda(x;\cc)&=&\Lambda_{0}(x)+\Lambda_{1}(x)\cc+\Lambda_{2}(x)\cc^{2}+\cdots~,\label{moi2}\\
j(x;\cc)&=&j_0(x)+j_1(x)\cc+j_2(x)\cc^{2}+\cdots~,\label{moi2-b}
\end{eqnarray}
where $\Lambda_{n}(x)$ and $j_n(x)$, $n=0,1,2,\cdots$, are
functions of $x$ only. The expansion of $j(x;\cc)$ follows
directly from those of $e^\lambda$  and $e^\nu$. It is obvious
that in the Newtonian limit $\cc \rightarrow 0$, $j(x )
\rightarrow j_0(x)=1$ and $\Lambda(x ) \rightarrow
\Lambda_0(x)=0$. On the other hand, integrating (\ref{moi1}) with
the regularity boundary condition of $\Lambda$ at $x=0$ leads to
\begin{equation}\label{moi3}
\frac{d\Lambda}{dx}
=-\frac{1}{x^{4}j}\int_{0}^{x}4x'^{3}\frac{dj}{dx'}(\Lambda-1) dx'
~.
\end{equation}
In consideration of the fact that
$dj/dx=j'_1(x)\cc+j'_2(x)\cc^{2}+\cdots~$, with $j'_n(x)\equiv
dj_n/dx$, we define $d \Lambda/dx$  as $\cc\Psi$, where $\Psi$ is
expandable in a regular power   series of compactness $\cc$ as:
\begin{equation}\label{Psi}
\Psi(x;\cc)=\Psi_{0}(x)+\Psi_{1}(x)\cc+\Psi_{2}(x)\cc^{2}+\cdots~,
\end{equation}
with $\Psi_{n} =d\Lambda_{n+1}/dx$ for $n\geq0$. For example, from
the expansions  (\ref{moi2}), (\ref{moi2-b}) and (\ref{moi3}) the
first two functions $\Psi_{0}(x)$ and $\Psi_{1}(x)$ are given
explicitly by the expressions:
\begin{eqnarray}
\Psi_{0}&=&-\displaystyle{\frac{1}{x^{4}j_{0}}}\int^{x}_{0}4x'^{3}j_{1}'(\Lambda_{0}-1)dx'~,\\
\Psi_{1}&=&\displaystyle{\frac{j_{1}}{x^{4}j_{0}^{2}}}\int^{x}_{0}4x'^{3}j_{1}'(\Lambda_{0}-1)dx'
-\frac{1}{x^{4}j_{0}}\int^{x}_{0}4x'^{3}\left[j_{2}'(\Lambda_{0}-1)+j_{1}'\Lambda_{1}\right]dx'~,
\end{eqnarray}
while the others can similarly be found. It can also be seen that
$\Psi_{n}$ is only related to $\Lambda_{i}$'s with $0 \le i\leq
n$.

On the other hand, the continuity boundary conditions of $\Lambda$
and $d\Lambda/dx$ at the stellar surface imply that at $x=1$
$\Lambda_{n+1}=-\Lambda_{n+1}'/3 $. Therefore, for $n \ge 0$
\begin{equation}
\Lambda_{n+1}(x)=\int_{1}^{x}\Psi_{n}(x')dx'-\Psi_{n}(1)/3~,~n=0,1,2,\ldots~.\label{moi4}
\end{equation}
As mentioned above, $\Psi_{n}$ depends only on $\Lambda_{n},
\Lambda_{n-1},\ldots, \Lambda_{0}$ and therefore $\Lambda_{n+1}$
can be determined from
$\Lambda_{0},\Lambda_{1},\ldots,\Lambda_{n}$ in an iterative way.
In particular, it can be shown from the recurrence relation
exhibited in (\ref{moi4}) that $\Psi_{n}$ and $\Lambda_{n}$ are
 $(2n+1)$-th and $2n$-th degree polynomials in $x$, respectively.

\begin{figure}[!htp]
\centering {\includegraphics[scale=0.4]{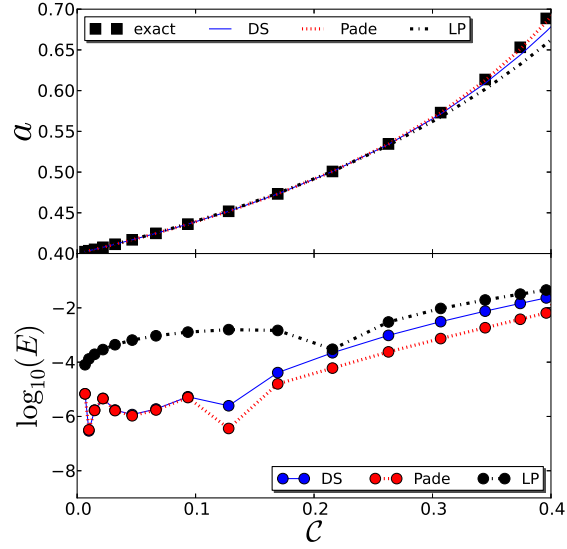}}
\caption{ Top panel: The geometric factor  $a\equiv I /(MR^2)$ is
plotted against compactness $\cc$. The exact numerical value is
denoted by solid squares. The approximate values of $a$ obtained
from DS with terms up to $\cc^6$ as shown in (\ref{moi7}), (2,2)
Pad\'{e} approximant (\ref{Pade_a}) and the LP empirical formula
(\ref{empirical}) are  denoted by the continuous solid, dotted and
dot-dashed
  curves, respectively. Bottom panel: The logarithm of the fractional deviations ($E$)
between the approximate values obtained from the above-mentioned
formulas and the exact numerical value of $a$ are plotted against
$\cc$. }\label{RMa1}
\end{figure}

Carrying out the above-mentioned steps recursively, we find in
turn the expansion of
 $\Lambda$,
\begin{equation}\label{moi5}
\Lambda(x;\cc)=\frac{2}{5}\left(5-3x^{2}\right)\cc
+\frac{1}{70}\left(21+126x^{2}-99x^{4}\right)\cc^{2}+\cdots~,
\end{equation}
which vanishes in the zero-$\cc$ limit as expected. The expansion
coefficients of $\Lambda(x;\cc)$ up to $\cc^6$-term are given in
Table~\ref{TSS4}. With the post-Minkowsian expansion for
$\Lambda(x;\cc)$, we can likewise find the expansion for the
geometric factor $a$:
\begin{equation}
a(\cc)=a_{0}+a_{1}\cc+a_{2}\cc^{2}+\cdots~,
\end{equation}
where the coefficients are given by $a_{n}=\hat{\Lambda}_{n+1}/2$.
The leading expansion (up to $\cc^6$-term) for  $a(\cc)$ is given
by
\begin{equation}
a(\cc)=\frac{2}{5}+\frac{12}{35}{\cal C}+\frac{212}{525}{\cal
C}^2+ \frac{632}{1155}{\cal C}^3+\frac{703744}{875875}{\cal
C}^4+\frac{251264}{202125}{\cal C}^5+
\frac{121542272}{60913125}{\cal C}^6+\cdots~.\label{moi7}
\end{equation}
In the zero-$\cc$ limit, $a=2/5$, which is  the exact result of a
Newtonian uniform sphere. The accuracy of (\ref{moi7}) for
relativistic incompressible stars is shown in Fig.~\ref{RMa1},
where the values of $a$ obtained from different schemes and the
logarithm of the corresponding fractional deviations, $E \equiv
|1-({\rm approximate~value}/{\rm exact~value})|$,  are plotted
against $\cc$ in the top and bottom panels, respectively. The
direct sum (DS) shown in (\ref{moi7}), including  terms up to
$\cc^6$, has an accuracy better than $0.01$ until $\cc$ is close
to $0.4$.

The accuracy of the expansion in  (\ref{moi7}) can be further
improved by constructing its Pad\'{e} approximants (see, e.g.,
\cite{Baker} for the theory and the construction of Pad\'{e}
approximants). In general, we can rewrite a $2m$-th
($m=1,2,3,\cdots$) order DS of a series $S$ such as (\ref{moi7})
into a $(m,m)$ diagonal Pad\'{e} approximant (denoted as
$\textrm{P}^{m}_{m}(S)$ hereafter), i.e., a rational function
whose numerator and denominator are $m$-th degree polynomials in
the expansion parameter ($\cc$ in our case). For example, the
(2,2) Pad\'{e} approximant of $a$ is given by
\begin{equation}\label{Pade_a}
\textrm{P}^{2}_{2}(a)=\frac{{\displaystyle
\frac{2}{5}-\frac{1154648}{2116205}{\cal
C}+\frac{24590348}{488843355}{\cal
C}^2}}{{\displaystyle1-\frac{940102}{423241}{\cal
C}+\frac{23746564}{23278255}{\cal C}^2}}  ~.
\end{equation}
As shown in Fig.~\ref{RMa1}, for $\cc >0.1$, the accuracy of
$\textrm{P}^{2}_{2}(a)$ is better than that of the DS containing
terms up to $\cc^6$, despite the fact that only terms up to
$\cc^4$ in the expansion (\ref{moi7}) are required in the
construction of $\textrm{P}^{2}_{2}(a)$. As long as $\cc <0.4$,
Eq.~(\ref{Pade_a})
 is accurate within $1\%$. For $\cc <0.1$,
the accuracies of these two formulas are almost identical.

We note that \citet{Lattimer:2001} have previously obtained an
empirical formula
\begin{eqnarray}
    a&=&\frac{2}{5(1-0.87\cc-0.3\cc^2)}\nonumber \\ &\approx& \frac{2}{5}(1 + 0.87 \cc + 1.0569
    \cc^2+ \cdots) \label{empirical}
\end{eqnarray}
   by fitting  the
numerical values of the moment of inertia of incompressible stars.
Comparing this empirical formula (termed as LP formula here) with
the
 analytic result in (\ref{moi7}), we see
that the former  is actually a good approximation of the latter at
low compactness because, as shown by the second equality in
(\ref{empirical}), to order $\cc^2$ these two expressions are
quite close. In a sense, the LP formula is a (0,2) Pad\'{e}
approximant for the geometric factor $a$. As shown in
Fig.~\ref{RMa1}, in the low-compactness regime, the fractional
deviation of the LP formula in (\ref{empirical}) is greater than
those of (\ref{moi7}) and (\ref{Pade_a}), reflecting the
significance of higher order terms in (\ref{moi7}). As $\cc $
grows larger $0.22$, the errors arising from the LP formula and
the DS (\ref{moi7}) are comparable, though the latter still
outperforms the former.  In general, the (2,2) Pad\'{e}
approximant in (\ref{Pade_a})  has the best accuracy among the
three approximations. This clearly demonstrates the advantage of
Pad\'{e} approximants.

\section{Tidal deformation}\label{Love}
In this section, we calculate the tidal Love number
(deformability) of incompressible stars and expand it in a series
of compactness. In a binary system with two compact stars, the
tidal field due to one compact star can induce quadrupole moments
on its companion. This effect can be quantified with either the
dimensionless tidal Love number $k_2$ or the tidal deformability
$\lambda$, which relates the applied tidal fields $E_{ij}$ to the
induced quadrupole moments $Q_{ij}$ through
\citep{Damour:09p084035,Lattimer_Love,Yagi:2013long,Yagi:2013,Yagi_14_PRD}
\begin{align}
Q_{ij}=-\frac{2k_2R^5}{3}E_{ij}\equiv -\lambda E_{ij}.\label{def}
\end{align}
The dimensionless  tidal deformability $\bar{\lambda}$ considered
in the I-Love-Q relations is defined by
$\bar{\lambda}\equiv\lambda/M^5=2k_2/(3\cc^5)$.

The calculations of the Love number $k_2$ are given in
\citep{Damour:09p084035,Hind2008}. In the present paper we mainly
follow the formulation established
 in \citep{Damour:09p084035}. With the conventions adopted in
 \citep{Price,Lindblom-1997}, in electric  tidal quadrupole deformation the metric perturbation
 $H=H_0=H_2$ satisfies the differential equation \citep{Damour:09p084035,Lattimer_Love}
\begin{align}
\label{Hfunction}
H''(r)+H'(r)\left[\frac{2}{r}+e^{\lambda(r)}\left(\frac{2m(r)}{r^2}+4\pi
r[p(r)-\rho(r)]\right)\right]+H(r)Q(r)=0,
\end{align}
where, in general,
\begin{align}
Q(r)=&4\pi
e^{\lambda(r)}\left[5\rho(r)+9p(r)+\frac{\rho(r)+p(r)}{c_s^2(r)}\right]-\frac{6e^{\lambda(r)}}{r^2}-[\nu'(r)]^2.
\end{align}
For incompressible stars, the density $\rho(r)=\rho_c$,  the sound
speed $c_s \equiv \sqrt{d p/d \rho}$ tends to infinity and hence
\begin{align}
Q(r)=&4\pi
e^{\lambda(r)}\left[5\rho(r)+9p(r)\right]-\frac{6e^{\lambda(r)}}{r^2}-[\nu'(r)]^2.
\end{align}
After solving $H$, the Love number $k_2$ can be found from the
following formula \citep{Damour:09p084035,Lattimer_Love}:
\begin{align}
\label{k2eq}
k_2({\cal C},y_R)=&\frac{8}{5}{\cal C}^5 (1-2{\cal C})^2[2{\cal C}(y_R-1)+2]\left\{2{\cal C}[4(y_R+1){\cal C}^4+(6y_R-4){\cal C}^3+(26-22y_R){\cal C}^2+3(5y_R-8)-3y_R+6]\right.\nonumber\\
&\left.+3(1-2{\cal C})^2[2{\cal C}(y_R-1)-y_R+2]\log(1-2{\cal C})
\right\}^{-1},
\end{align}
where
\begin{eqnarray}
 y_R &\equiv& \left(\frac{r}{H}\frac{d H}{d r}\right)_{r=R^+}\nonumber \\
  &= &\left(\frac{r}{H}\frac{d H}{d r}-\frac{4 \pi R^3
\rho}{M}\right)_{r=R^-}.\label{bc}
\end{eqnarray}
It is readily shown from (\ref{Hfunction}) that inside the star
the logarithmic derivative of $H$, $y \equiv rH'(r)/H(r)$,  is
governed by \citep{Lattimer_Love}
\begin{align}
\label{ylove} r y'(r)+y(r)^2&+y(r)e^{\lambda(r)}\left\{1+4\pi
r^2[p(r)-\rho(r)]\right\}+r^2Q(r)=0,
\end{align}
and the appropriate boundary condition at the center is $y(0)=2$.
Once $y(r)$ and in turn $y_R$ are obtained, the Love number $k_2$
can be found from (\ref{k2eq}).

Similar to the series expansion of $I$, we assume that $y$ can be
expanded as a power series in ${\cal C}$
\begin{align}
\label{yexp} y(x)=y_0(x)+y_1(x){\cal C}+y_2(x){\cal
C}^2+y_3(x){\cal C}^3+\cdots~,
\end{align}
where $y_n(x)$, $n=0,1,2,\cdots$, are functions of $x$ only. We
substitute \eqref{yexp} and the series expansions of $p$, $e^\nu$
and $e^\lambda$
 (see (\ref{expp}), (\ref{expnu})  and  (\ref{explamb}))  into
\eqref{ylove} and solve the resultant equation order by order,
leading to a set of first-order ordinary differential equations:
\begin{equation}
\label{firsty} x y_0'(x)+y_0(x)^2+y_0(x)-6=0,
\end{equation}
\begin{equation}
x y_1'(x)+[1+2y_0(x)]y_1(x)-x^2y_0(x)+3x^2=0,
\end{equation}
and etc for $0 \le x <1$. As  $y_n(x)$ do not depend on ${\cal
C}$, it follows directly from the boundary condition $y(0)=2$ that
$y_0(0)=2$ and $y_n(0)=0$ for $n>0$. This leads to
\begin{eqnarray}
y_0(x)&=&2;\;\;\;\\y_1(x)&=&-\frac{1}{7}x^2,
\end{eqnarray}
and higher-order expansion can be obtained recursively. For
reference the explicit expressions of $y_n$ up to $n=6$ are
presented in Table \ref{yRseries}.

\begin{figure}[!htp]
\includegraphics[scale=0.4]{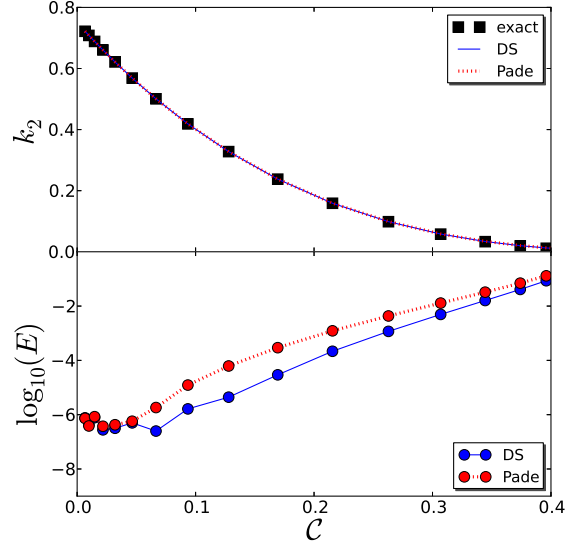}
\caption{Top panel: The tidal Love number $k_2$ is plotted against
compactness ${\cal C}$. Sold squares represent exact numerical
data. The continuous solid curve and dotted curve denote
approximate values of $k_2$ obtained from DS in  (\ref{k2}) with
terms up to $\cc^6$ and $(2,2)$ Pad\'{e} approximant
(\ref{k2pade}), respectively. Bottom panel: The logarithms of the
fractional deviations for the two  above-mentioned approximations
from the exact value of $k_2$ are shown versus $\cc$.}
\label{incom_k2}
\end{figure}

Upon substitution of the so obtained $y_n$  into (\ref{yexp}) and
by (\ref{k2eq}) and (\ref{bc}), we can find the post-Minkowsian
expansion for the Love number of incompressible stars:
\begin{align}\label{k2}
k_2&=\left(1-2{\cal C}\right)^2\left(\frac{3}{4}-\frac{9}{7}{\cal
C}+\frac{121}{294}{\cal C}^2 -\frac{479}{11319}{\cal
C}^3-\frac{196375}{1030029}{\cal
C}^4-\frac{10670812}{21630609}{\cal C}^5
-\frac{32621700682}{28314467181}{\cal C}^6+\cdots\right),
\end{align}
and its (2,2) Pade approximation
\begin{align}\label{k2pade}
{\rm P}^2_2(k_2)&=\left(1-2{\cal C}\right)^2
\frac{\frac{3}{4}+\frac{1213215}{2842294}{\cal
C}-\frac{656811130}{328284957}{\cal
C}^2}{1+\frac{3245062}{1421147}{\cal
C}+\frac{76383026}{109428319}{\cal C}^2}~.
\end{align}
In these two expressions for $k_2$, (\ref{k2}) and (\ref{k2pade}),
we have deliberately kept the factor $(1-2\cc)^2$ intact to
manifest the black-hole limit as suggested in
\citep{Damour:09p084035}. Figure~\ref{incom_k2} compares the exact
numerical value of the Love number with the approximate values
obtained from (\ref{k2}) and (\ref{k2pade}). In general, the
agreement between the exact and the two approximate values is
good, especially for small ${\cal C}$. As long as ${\cal C} <
0.3$, the fractional deviations of the two approximations  in
(\ref{k2}) and (\ref{k2pade}) are less than $0.01$, while the
former is slightly smaller than latter. It should be noted that
our main objective here is to derive the I-Love relation for
realistic stars, whose compactnesses are in most cases less than
0.3. Therefore, both Eqs.~(\ref{k2}) and (\ref{k2pade}) are
accurate enough for the following discussion.

The expansion of the dimensionless tidal deformability of
incompressible stars follows directly from (\ref{k2}),
\begin{equation}\label{lambdaC}
\bar{\lambda}=\frac{1}{2 {\cal C}^5}-\frac{20}{7 {\cal C}^4}+\frac{2515}{441
   {\cal C}^3}-\frac{51550}{11319 {\cal C}^2}+\frac{3347350}{3090087
   {\cal C}}+\frac{4326424}{64891827}+ \frac{368458100}{9438155727}{\cal C}+\cdots.
\end{equation}
It is clearly seen from the above series that $\bar{\lambda}$
diverges as $1/(2 {\cal C}^{5})$ in the Newtonian limit. Besides,
by inverting (\ref{lambdaC}), we can express the compactness in
terms of a power series in $\zeta \equiv (2\bar{
\lambda})^{-1/5}$:
\begin{equation}\label{C_zeta}
\cc=\zeta-\frac{8 }{7}\zeta^2+\frac{430 }{441}\zeta^3-\frac{24020
}{33957}\zeta^4+\frac{1338940 }{3090087}\zeta^5-\frac{207659912
}{973377405}\zeta^6+\frac{130446932288}{1638194172615}\zeta^7
+\cdots~,
\end{equation}
or equivalently
\begin{equation}\label{Clambda}
{\cal C}=\frac{0.8706}{\bar{\lambda}^{1/5}}
-\frac{0.8661}{\bar{\lambda}^{2/5}}+\frac{0.6433}{\bar{\lambda}^{3/5}}
-\frac{0.4063}{\bar{\lambda}^{4/5}}+\frac{0.2167}{\bar{\lambda}}-\frac{0.09286}{\bar{\lambda}^{6/5}}
+\frac{0.03017}{\bar{\lambda}^{7/5}}+\cdots~.
\end{equation}
Eq.~(\ref{C_zeta}) or (\ref{Clambda}) is useful in the derivation
of the I-Love relation.
\section{I-Love relation for compact stars}\label{I_Love}
\citet{Yagi:2013} pointed out that there exists  an almost
EOS-independent relationship between the moment of inertia and the
tidal Love number (deformability) of compact stars, which is
universal to within $1 \%$. Their empirical fitting curve for the
I-Love relation (referred to as YY formula here) is
\begin{equation}
\label{eqYY} \ln
\bar{I}=1.47+0.0817\ln\bar{\lambda}+0.0149(\ln\bar{\lambda})^2+2.87\times
10^{-4}(\ln\bar{\lambda})^3-3.64\times 10^{-5}(\ln
\bar{\lambda})^4,
\end{equation}
with $\bar{I} \equiv I/M^3=a/\cc^{2}$ being the dimensionless
moment of inertia. The first (from top to bottom) panel of
Fig.~\ref{effcom_lambda_yycom} shows a plot of $\log_{10}\bar{I}$
versus $\log_{10}\bar{\lambda}$ for realistic NSs with seven
realistic EOSs (including APR \citep{APR}, AU \citep{WFF}, BBB2
\citep{BBB2}, FPS \cite{FPS,Lorenz:93p379}, SLy4 \citep{DH2001},
UU \citep{WFF}, WS \citep{Lorenz:93p379,WFF},  one QS
characterized by the MIT bag model \citep{MITBM,Witten} and
incompressible stars. It is seen that the YY formula is a good
approximation as long as $\log_{10}\bar{\lambda} < 4$. However, as
can be observed from the second panel of
Fig.~\ref{effcom_lambda_yycom}, where the fractional deviation of
the the YY formula (denoted by $E_{YY}$)
 is shown, the accuracy of the YY formula significantly
worsens as $\log_{10}\bar{\lambda}$ grows beyond 4. In other
words, the YY formula (\ref{eqYY}) does not cover the case of
Newtonian stars.

Combining the series expansions   (\ref{moi7}) and (\ref{C_zeta})
(or, equivalently, (\ref{Clambda})), we can find an analytic
I-Love relation for incompressible stars up to any desirable
powers in $\bar{\lambda}^{-1/5}$. For example, the explicit
expression for $\bar{I}$ up to $\zeta^4$ (or
$\bar{\lambda}^{-4/5}$) is given by
\begin{eqnarray}
\label{incomilove_ds} \bar{I}&=& \frac{2}{5 \zeta^2}+\frac{44}{35
\zeta}+\frac{17452}{11025}+\frac{31936
}{33957}\zeta+\frac{21242792 }{105343875}\zeta^2-\frac{990746384
}{24334435125}\zeta^3-\frac{59041871509888}{1433419901038125}\zeta^4
+\cdots~,\nonumber  \\
&=&
\bar{\lambda}^{2/5}\left(0.5278+\frac{1.444}{\bar{\lambda}^{1/5}}+\frac{1.583}{\bar{\lambda}^{2/5}}
+\frac{0.8187}{{\bar{\lambda}^{3/5}}}
+\frac{0.1528}{{\bar{\lambda}^{4/5}}}-\frac{0.02686}{{\bar{\lambda}}}-\frac{0.02366}{{\bar{\lambda}^{6/5}}}
+\cdots \right)~.
\end{eqnarray}
From the above DS formula, we can construct   the $(m,m)$
 diagonal Pad\'{e} approximant ($m=1,2,,\cdots$) for $\bar{I}
 \zeta^2$
(or $\bar{I}/\bar{\lambda}^{2/5}$) by considering $\zeta$ (or
$\bar{\lambda}^{-1/5}$) as the expansion parameter \cite{Baker}.
The formula for $\bar{I}$ resulting from the (2,2) Pad\'{e}
approximant, which is constructed from the leading five terms of
the DS in (\ref{incomilove_ds}), is explicitly given
 by
\begin{eqnarray}\label{incomlovei_Pade}
\bar{I}&=&\frac{\frac{2}{5}+\frac{44471656496
}{50319113845}\zeta+\frac{2891441990432
}{4981592270655}\zeta^2}{\zeta^2\left(1-\frac{9393329026
}{10063822769}\zeta+\frac{4965191579746
}{11623715298195}\zeta^2\right)}~,
\nonumber \\
&=&\bar{\lambda}^{2/5}\left(\frac{0.5278+1.015\bar{\lambda}^{-1/5}+0.5804\bar{\lambda}^{-2/5}}
{1-0.8126\bar{\lambda}^{-1/5}+0.3237\bar{\lambda}^{-2/5}}\right)~.
\end{eqnarray}

Although both (\ref{incomilove_ds}) and (\ref{incomlovei_Pade})
are intended for incompressible stars, as suggested in
\cite{ILoveQ_1} and shown in Fig.~\ref{effcom_lambda_yycom}, they
are indeed good approximation for realistic NSs and QSs. In
particular, the fractional deviations, $E_{DS}$ and $E_{Pade}$,
between $\bar{I}$ of various realistic NSs and QSs and the
approximate values $\bar{I}$ of incompressible stars given by
(\ref{incomilove_ds}) and (\ref{incomlovei_Pade}), respectively,
are plotted against $\log_{10}\bar{\lambda}$ in the third and
fourth panels of Fig.~\ref{effcom_lambda_yycom}. It is clearly
seen that $E_{DS}$ and $E_{Pade}$ are generally comparable to each
other. They are less than 0.02 for all realistic stars. For stars
close to the maximum mass limit, $E_{DS}$ is still within $0.01$
and smaller than $E_{Pade}$. On the other hand, the two deviations
are very small towards the Newtonian limit. Hence, both
(\ref{incomilove_ds}) and (\ref{incomlovei_Pade}) suffice to
express the universal I-Love relation for realistic NSs,  QSs and
incompressible stars as well with good accuracy from the Newtonian
limit to the maximum mass limit.


\section{Conclusion and discussion}\label{Conclusion}
In this paper, we derive from first principle the post-Minkowskian
expansions for the moment of inertia and the Love number (or tidal
deformability) of incompressible stars, expressing both physical
quantities in terms of power series or Pad\'{e} approximants of
the compactness of the star. To our knowledge, such analytic
expansions have not been available previously (see, e.g., the
remark in \citep{Lattimer:2001}). The high accuracy of these
formulas readily guarantees that they are useful in their own
right. Due to the recursive nature of the expansion method
employed in our derivation, the formulas for both quantities can
be straightforwardly extended to higher orders should better
precision for stars with large compactness is needed.

\begin{figure}[!htp]
\includegraphics[scale=0.4]{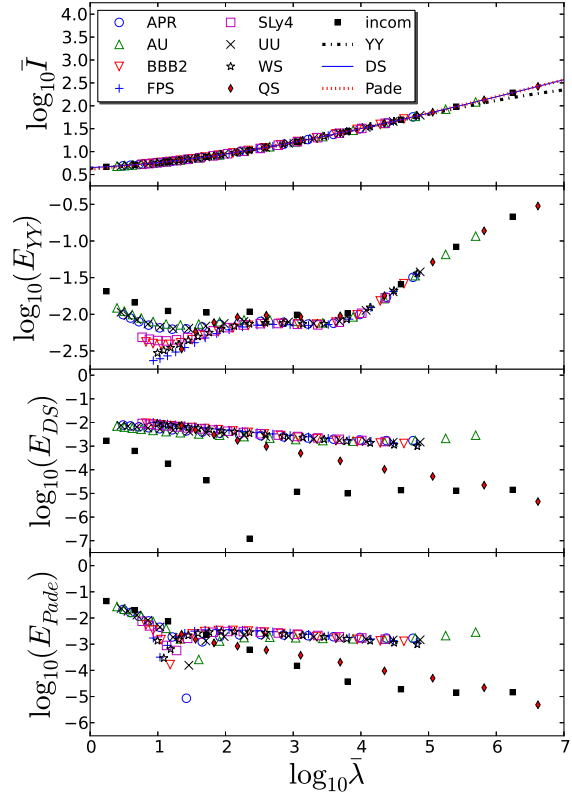}
\caption{First (from top to bottom) panel: $\log_{10}\bar{I}$
versus $\log_{10}\bar{\lambda}$. Exact data of NSs with seven
realistic EOSs (including APR, AU, BBB2, FPS, SLy4, UU, WS),  one
QS and incompressible stars are shown. Besides, approximate values
obtained from YY formula (\ref{eqYY}), sixth-order DS formula
(\ref{incomilove_ds}) and $(2,2)$ Pad\'{e} approximant
(\ref{incomlovei_Pade}), respectively denoted by the dot-dashed,
continuous  and dotted curves, are also included.
Second/Third/Fourth panel: The logarithm of the fractional
deviation $E_{YY}$/$E_{DS}$/$E_{Pade}$ between the YY/sixth-order
DS/$(2,2)$ Pad\'{e} approximation of $\bar{I}$  and the exact
value of $\bar{I}$ is shown against $\log_{10}\bar{\lambda}$.}
\label{effcom_lambda_yycom}
\end{figure}

Furthermore, by eliminating the compactness from these two
formulas, we obtain explicit equations, (\ref{incomilove_ds})  and
(\ref{incomlovei_Pade}), directly connecting the moment of inertia
and the Love number (or tidal deformability) of incompressible
stars. Most importantly, Eqs.~(\ref{incomilove_ds})  and
(\ref{incomlovei_Pade}) are accurate enough to predict the
universal  behavior of realistic NSs and QSs to within $2 \%$
level (see Fig.~\ref{effcom_lambda_yycom}). In fact, the agreement
is much better than $1 \%$ as long as $\bar{\lambda}>10$ (i.e.,
not too dense stars). In comparison with the existing empirical
formula, the present one  has a broader range of applicability and
is physically more transparent. For example, in the Newtonian
limit, $\bar{\lambda} \rightarrow \infty$, it follows directly
from (\ref{incomlovei_Pade}) that
$\bar{I}=2^{7/5}\bar{\lambda}^{2/5}/5 $, which agrees nicely with
the Newtonian formula for incompressible stars obtained by
\citet{Yagi:2013long}.

Our finding reported here also strongly supports the claim of
\citet{ILoveQ_1}  that the I-Love-Q universal relations of
realistic NSs and QSs follow closely  those of incompressible
stars. As pointed out in \citep{ILoveQ_1}, the physical origins of
the I-Love-Q universal relations \citep{Yagi:2013long,Yagi:2013}
are the high stiffness of dense nuclear/quark matter and, in
addition, the I-Love-Q relations are also insensitive to variation
in stiffness around the incompressible limit. The good agreement
between the theoretical I-Love formula and realistic data shown in
Fig.~\ref{effcom_lambda_yycom} clearly lends support to the views
proposed in \citep{ILoveQ_1}.

 To further justify theoretically  the claim that the
I-Love-Q relations are insensitive to variation in stiffness
around the incompressible limit \citep{ILoveQ_1}, explicit I-Love
formulas for stars characterized by different stiffness (e.g.
polytropic stars) should be sought. However, the derivation of the
I-Love formula reported here crucially relies on the availability
of the analytical solutions of the TOV equations for
incompressible stars. To our knowledge, exact solutions  for TOV
equations of  polytropic stars have not yet been derived. It will
be helpful if such solutions can be worked out.

Lastly, a remark about the validity of (\ref{incomilove_ds})  and
(\ref{incomlovei_Pade}) is in order. For a realistic NS near its
low mass limit (typically, $\bar{\lambda} > 10^8$) , a significant
portion of its mass content is comprised of soft nuclear matter
with polytropic index greater than unity. Hence, its behavior can
no longer be approximated by the incompressible limit
\citep{ILoveQ_1}. The accuracy of (\ref{incomilove_ds})  and
(\ref{incomlovei_Pade}) is expected to worsen in this situation.
However, NSs with such low compactness are not particularly
interesting in astrophysics.

\begin{acknowledgments}
 We thank L.-M.~Lin,  Y.H.~Sham  and H.K.~Lau for helpful discussions.
 We are also grateful to Lin and Sham for providing us some
 numerical data presented here.
 \end{acknowledgments}


\newcommand{\noopsort}[1]{} \newcommand{\printfirst}[2]{#1}
  \newcommand{\singleletter}[1]{#1} \newcommand{\switchargs}[2]{#2#1}

\clearpage

\begin{table}[h]
\caption{The coefficient of the $x^j$-term in the polynomials
$\bar{p}_n(x) \equiv {p}_n/\rho_c$ for $0 \le n \le 6$. There is
no odd power term and the coefficient vanishes if $j >2n$.
}\label{TSS2} \centering
\begin{tabular}{c|cccccccc}
&~~~$x^0$~~~&$~~~x^2~~~$&~~~$x^4~~~$&~~~$x^6$~~~&~~~$x^8$~~~&$~~x^{10}$~~&$~~x^{12}$~~\\
\hline
$\bar{p}_0$& 0 & 0 & 0 & 0 & 0 & 0 & 0 \\
$\bar{p}_1$& $\frac{1}{2}$ & $-\frac{1}{2}$ & 0 & 0 & 0 & 0 & 0 \\
$\bar{p}_2$& 1 & $-1$ & 0 & 0 & 0 & 0 & 0 \\
$\bar{p}_3$& $\frac{17}{8}$ & $-\frac{19}{8}$ & $\frac{3}{8}$ & $-\frac{1}{8}$ & 0 & 0 & 0 \\
$\bar{p}_4$& $\frac{37}{8}$ & $-\frac{23}{4}$ & $\frac{3}{2}$ & $-\frac{1}{4}$ & $-\frac{1}{8}$ & 0 & 0 \\
$\bar{p}_5$& $\frac{163}{16}$ & $-\frac{223}{16}$ & $\frac{39}{8}$ & $-\frac{7}{8}$ & $-\frac{1}{16}$ & $-\frac{3}{16}$ & 0 \\
$\bar{p}_6$& $\frac{723}{32}$ & $-\frac{539}{16}$ & $\frac{465}{32}$ & $-\frac{25}{8}$ & $\frac{5}{32}$ & $-\frac{3}{16}$ & $-\frac{9}{32}$ \\


\end{tabular}
\end{table}
\begin{table}[h]
\caption{The coefficient of the $x^j$-term in the polynomials
$(e^{\nu})_{n}(x)$
  for $0 \le
n \le 6$. There is no odd power term and the coefficient vanishes
if $j >2n$. } \label{TSS3} \centering
\begin{tabular}{c|cccccccc}
&$~~~x^0~~~$&$~~~x^2~~~$&$~~~x^4~~~$&$~~~x^6~~~$&$~~~x^8~~~$&$~~x^{10}~~$&$~~x^{12}~~$\\
\hline
$(e^{\nu})_{0}$& 1 & 0 & 0 & 0 & 0 & 0 & 0 \\
$(e^{\nu})_{1}$& $-3$ & 1 & 0 & 0 & 0 & 0 & 0 \\
$(e^{\nu})_{2}$& $\frac{3}{4}$ & $-\frac{3}{2}$ & $\frac{3}{4}$ & 0 & 0 & 0 & 0 \\
$(e^{\nu})_{3}$& $\frac{3}{4}$ & $-\frac{3}{4}$ & $-\frac{3}{4}$ & $\frac{3}{4}$ & 0 & 0 & 0 \\
$(e^{\nu})_{4}$& $\frac{15}{16}$ & $-\frac{3}{4}$ & $-\frac{3}{8}$ & $-\frac{3}{4}$ & $\frac{15}{16}$ & 0 & 0 \\
$(e^{\nu})_{5}$& $\frac{1171}{80}$ & $-\frac{15}{16}$ & $-\frac{3}{8}$ & $-\frac{3}{8}$ & $-\frac{15}{16}$ & $\frac{21}{16}$ & 0 \\
$(e^{\nu})_{6}$& $-\frac{2569}{240}$ & $\frac{961}{80}$ &
$-\frac{15}{32}$ & $-\frac{3}{8}$ & $-\frac{15}{32}$ &
$-\frac{21}{16}$ & $\frac{63}{32}$
\\

\end{tabular}

\end{table}


\begin{table}[ht]
\caption{The coefficient of the $x^j$-term in the polynomials
$\Lambda_{n}(x)$ for $0 \le n \le 6$. There is no odd power term
and the coefficient vanishes if $j >2n$. }\label{TSS4} \centering
\begin{tabular}{c|cccccccc}
$~~~$&$~~~x^0~~~$&$~~~x^2~~~$&$~~~x^4~~~$&$~~~x^6~~~$&$~~~x^8~~~$&$~~x^{10}~~$&$~~x^{12}~~$\\
\hline
$\Lambda_0$&0 & 0 & 0 & 0 & 0 & 0 & 0 \\
$\Lambda_1$&2 & $-\frac{6}{5}$ & 0 & 0 & 0 & 0 & 0 \\
$\Lambda_2$& $\frac{3}{10}$ & $\frac{9}{5}$ & $-\frac{99}{70}$ & 0 & 0 & 0 & 0 \\
$\Lambda_3$& $\frac{2}{7}$ & $\frac{9}{25}$ & $\frac{72}{35}$ & $-\frac{199}{105}$ & 0 & 0 & 0 \\
$\Lambda_4$& $\frac{2869}{8400}$ & $\frac{261}{700}$ & $\frac{549}{1400}$ & $\frac{11}{4}$ & $-\frac{3403}{1232}$ & 0 & 0 \\
$\Lambda_5$& $\frac{1753}{3850}$ & $\frac{3439}{7000}$ & $\frac{927}{2450}$ & $\frac{373}{700}$ & $\frac{1541}{385}$ & $-\frac{851547}{200200}$ & 0 \\
$\Lambda_6$& $\frac{54247603}{84084000}$ & $\frac{113063}{154000}$ & $\frac{88737}{196000}$ & $\frac{5169}{9800}$ & $\frac{9557}{12320}$ &  $\frac{493299}{80080}$ & $-\frac{11684429}{1716000}$\\
\end{tabular}
\end{table}



\begin{table}[h]
\caption{The coefficient of the $x^j$-term in the polynomials
$y_{n}(x)$
  for $0 \le
n \le 6$. There is no odd power term and the coefficient vanishes
if $j >2n$.}\label{yRseries} \centering
\begin{tabular}{c|cccccccc}
&$~~~x^0~~~$&$~~~x^2~~~$&$~~~x^4~~~$&$~~~x^6~~~$&$~~~x^8~~~$&$~~~x^{10}~~~$&$~~~x^{12}~~~$\\
\hline
$y_0$&2 & 0 & 0 & 0 & 0 & 0 & 0 \\
$y_1$& 0 & $-\frac{1}{7}$ & 0 & 0 & 0 & 0 & 0 \\
$y_2$& 0 & $-\frac{33}{14}$ & $\frac{599}{294}$ & 0 & 0 & 0 & 0 \\
$y_3$& 0 & $-\frac{33}{7}$ & $\frac{50}{49}$ & $\frac{12046}{3773}$ & 0 & 0 & 0 \\
$y_4$& 0 & $-\frac{561}{56}$ & $\frac{1643}{392}$ & $-\frac{18561}{30184}$ & $\frac{146821487}{24720696}$ & 0 & 0 \\
$y_5$& 0 & $-\frac{1221}{56}$ & $\frac{634}{49}$ & $-\frac{519}{343}$ & $-\frac{468151}{2060058}$ & $\frac{209428151}{19227208}$ & 0 \\
$y_6$ &0 & $-\frac{5379}{112}$ & $\frac{28475}{784}$ & $-\frac{3345}{616}$ & $\frac{10010095}{8240232}$ & $-\frac{85525955}{115363248}$ & $\frac{27526499425793}{1359094424688}$\\
\end{tabular}
\end{table}
\clearpage

\end{document}